\documentclass[]{interact}
\usepackage{algorithm}
\usepackage{algpseudocode}
\usepackage{amssymb}
\usepackage{hyperref}
\usepackage{wrapfig}
\usepackage{subcaption}
\usepackage{tikz}
\usepackage{xfrac}
\hypersetup{
    colorlinks=true,
    linkcolor=blue,
    filecolor=blue,      
    urlcolor=blue,
    citecolor=blue,
}

\usepackage{epstopdf}% To incorporate .eps illustrations using PDFLaTeX, etc.

\usepackage{natbib}% Citation support using natbib.sty
\bibpunct[, ]{(}{)}{;}{a}{}{,}% Citation support using natbib.sty
\renewcommand\bibfont{\fontsize{10}{12}\selectfont}% Bibliography support using natbib.sty

\theoremstyle{plain}% Theorem-like structures provided by amsthm.sty

\theoremstyle{definition}

\theoremstyle{remark}

\usepackage{changes}
\makeatletter
\@namedef{Changes@AuthorColor}{red}
\colorlet{Changes@Color}{red}

\usepackage{lineno}
\begin{document}

\articletype{}

\title{Optimal frequency setting of metro services in the age of COVID-19 distancing measures}

\author{
\name{Konstantinos Gkiotsalitis\textsuperscript{a}\thanks{CONTACT: K. Gkiotsalitis. Email: k.gkiotsalitis@utwente.nl; ORCID: https://orcid.org/0000-0002-3009-1527} and Oded Cats\textsuperscript{b}}
\affil{\textsuperscript{a}University of Twente, Horst-Ring Z-222, P.O. Box 217, 7500 AE Enschede, The Netherlands;
\textsuperscript{b}Delft University of Technology, P.O. Box 5048, 2600 GA Delft, The Netherlands}
}

\maketitle

\begin{abstract}
Public transport is one of the most disrupted sectors of the COVID-19 pandemic with reported ridership drops up to 90\% in majorly affected countries. As many government authorities strive to partially resume activities, public transport operators are in an urgent need for models that can evaluate the impact of different social distancing policies on operational and passenger-related costs. In this study, we introduce a mixed-integer quadratic programming model for the redesign of public transport services considering the operational, passenger, and revenue loss-related costs by evaluating the effects of different social distancing policies. Our model is applied at the metro network of Washington D.C. and provides optimal redistribution of vehicles across lines for different social distancing scenarios. This model can be used as a decision support tool by other policymakers and public transport operators that are in need of evaluating the costs related to the implementation of different social distancing policies.
\end{abstract}


\section*{Word Count: 6,320; Tables: 6 and Figures: 6}

\section{Introduction}\label{sec1}
Coronaviruses are highly contagious respiratory pathogens leading public authorities to implement strict distancing measures in closed spaces, such as schools, shops, working places, and public transport. The COVID-19 epidemic was first reported to the World Health Organization (WHO) on December 31, 2019 and it was recognized as a pandemic on March 11, 2020 \citep{who2020}. Public transport is one of the most disrupted sectors of the COVID-19 pandemic with early estimates suggesting that the drop has been as much as 80-90\% in major cities in China, Iran and the U.S., and as much as 70\% for some operators in the U.K \citep{UITP2020}.

Several public transport operators have reduced their service span (e.g. have canceled night services), have reduced service frequencies, and have closed selected public transport stations. For example, Transport for London (TfL) has suspended the night tube service and closed 40 metro stations that do not interchange with other lines \citep{tfl2020}. Similarly, the Washington Metropolitan Area Transit Authority (WMATA) closed 19 metro stations out of 91 and requested all passengers to wear a cloth face-covering \citep{wmata2020customers}. In addition, it has reduced the service frequencies from 10 trains per hour to 3 or 4 trains per hour during peak hours and has imposed early rail closures at 9pm \citep{wmata2020covid}.

Government authorities propose social distancing rules ranging from 1- to 2-meter distancing in closed spaces because large droplets do not travel further than 2 meters \citep{bahl2020airborne,jarvis2020quantifying}. Consequently, public transport service providers have to re-design their services while considering the local distancing regulations. This calls for the development of novel methods for managing the limited capacity available and allocating resources accordingly so as to minimize the ramifications of the newly imposed constraints driven by public health considerations. 

The limited capacity implied by the corona-related distancing measures requires the reallocation of public transport resources so as to cater most efficiently and effectively for the prevailing demand patterns while maintaining the functionality of the public transport system. Notwithstanding, even when all available resources (vehicles, drivers) are deployed, it is expected that not all passenger demand can be absorbed along the busiest service segments. Determining the optimal reallocation of service resources, which also implies the determination of which demand segments may not be satisfied, is not trivial. Most public transport networks are denser in their high-demand core and become thinner as they branch out. During peak periods, passenger load levels in the core of the network are often such that it is not possible to safely transport all passengers. Moreover, passengers boarding at the edges of the central area may not be able to board as the occupancy level is already approaching the new corona-era capacity standard, leaving many stranded passengers. Should one cater for the long-distance low-volume travel from the branches to the core and vice-versa or for the short-distance high-volume demand within the network core?  

Existing public transport network design \citep {mandl1980evaluation,ceder1986bus,pattnaik1998urban,borndorfer2007column,szeto2011simultaneous,ul2018comparison} and frequency setting \citep{gkiotsalitis2018reliable,sun2019optimal,gkiotsalitis2019cost} methods cannot answer the above-mentioned research question because they consider only the trade-off between operational costs (e.g., running costs, in-vehicle occupancy levels) and passenger-related costs (e.g., waiting times at stations, total trip travel times). Thus, these models do not account for the implications of the implementation of social distancing measures and cannot, therefore, support public transport service providers in the planning of their services in the era of COVID-19. 

To this end, we develop a method that supports the re-design of mass transit services in the context of complying with COVID-19 distancing measures. More specifically, we formulate and solve a network-wide model that can set the optimal frequencies of services lines under different social distancing scenarios. The proposed model extends on the classic trade-off between operational-related and passenger-related costs by considering the revenue losses associated with the unaccommodated passenger demand when complying with the distancing measures. Our network-wide frequency setting model can be used by public transport operators that seek to re-design their services under different distancing scenarios and investigate their performance in terms of passenger waiting time costs, operational costs, in-vehicle occupancy levels, violations of distancing standards, and revenue losses due to denied boarding. 

The remainder of this study is structured as follows: in section \ref{sec2} we review frequency setting models and introduce the well-established model of \citet{furth1981setting} which is adopted and adapted in this study. Section \ref{sec3} presents our network-wide frequency setting model that considers the impacts of distancing measures. Our model is formulated as a mixed-integer nonlinear program (MINLP) and it is reformulated to an easier-to-solve mixed-integer quadratic program (MIQP) that can be applied to different case studies. In our case study (section \ref{sec4}), we apply our model to compute the optimal service frequencies of the Washington D.C. metro lines under different distancing policies (no distancing, 1-meter distancing, 1.5-meter distancing, 2-meter distancing). In addition, we investigate the impact of different distancing policies on operational costs, passenger-related costs, vehicle occupancy levels, and revenue losses due to denied boarding. This is instrumental for public transport service providers that need to plan their operations while taking into consideration public health risks and the operational/passenger-related costs, as elaborated in our discussion (section \ref{sec5}).

\section{Literature review and baseline frequency setting model}\label{sec2}
Frequency setting models determine the allocation of the available fleet to different line services by considering the trade-off between productivity and operational costs \citep{ibarra2015planning}. Earlier works were limited to determining the frequency of a single line at a time \citep{furth1981setting,ceder1984bus,ceder2002urban}. \citet{ceder1984bus} proposed closed-form expressions that do not need to solve complex mathematical programs when determining the frequency of a single line (namely, the maximum load and the load profile methods). 

In the last decade, a series of models have been proposed for setting service frequencies network-wide by determining the optimal resource allocation subject to limited resources. \citet{yu2010genetic} proposed a bi-level programming model for the frequency setting problem which determines the optimal frequencies by minimizing the total travel time of passengers subject to overall fleet size limitations. The optimal frequency setting and allocation of a mixed-fleet was considered by \citet{cats2019frequency} and \citet{dell2012optimizing}. Using meta-heuristics, the frequency setting problem was integrated with the route design by \citet{szeto2011simultaneous} and \citet{arbex2015efficient}.
\citet{cipriani2012transit} also addressed the frequency setting problem as an exercise of balancing the passenger demand with the available supply.  \citet{verbas2013optimal,verbas2015stretching,verbas2015exploring,gkiotsalitis2017exact} and \citet{gkiotsalitis2019cost} developed frequency setting models that consider flexible virtual lines (e.g., short-turning/interlining lines) to exploit the available vehicle/driver resources as much as possible. The works of \citet{delle1998service} and \citet{cortes2011integrating} also focus on generating short-turning lines to serve the excessive demand at crowded line segments.

In this study we adopt the model formulation proposed by \citet{furth1981setting} as a baseline. This formulation is limited to a single line and considers the operational costs expressed by the number of vehicles required and the passenger-related costs expressed by the total passenger waiting times.

To describe the approach of \citet{furth1981setting}, let us consider a time period of one hour and a number of available vehicles, $N$. Let also $x$ be the number of vehicles allocated to the service line and $f$ the resulting frequency. Finally, let $T$ be the round-trip travel time and $B_{sy}$ the expected passenger demand between any origin-destination pair within the 1-hour time period (note that $B_{sy}$ corresponds to the hourly passenger arrival rate). Then, the optimal service frequency is determined by the model of \citet{furth1981setting} by solving the following mixed-integer nonlinear program (MINLP):

\begin{align}
(\text{MINLP})\quad & \min_{x,f} & W\cdot x+\sum\limits_{s=1}^{|\mathcal{S}|-1}\sum\limits_{y=s+1}^{|S|}B_{sy}\frac{1}{f}\\
&\text{subject to}&x\leq N\label{eq2}\\
&& f\leq f_{max}\label{eq3}\\
&& f\leq \frac{x}{T}\label{eq4}\\
&& x\in\mathbb{Z}_{\geq 0}\label{eq5}\\
&& f\in \mathbb{R}_{\geq 1}\label{eq6}
\end{align}

The objective function of the above program minimizes two components: (i) $Wx$ which is the operational cost expressed by the number of assigned vehicles $x$ multiplied by a weight factor $W$ associated with the cost of operating an extra vehicle, and (ii) $\sum\limits_{s=1}^{|\mathcal{S}|-1}\sum\limits_{y=s+1}^{|S|}B_{sy}\frac{1}{f}$ which is the passenger-related cost expressed by the number of passengers traveling between each origin-destination pair $sy$ multiplied by the inverse of the service frequency $f$ (note that the higher the  service frequency, the less passengers $B_{sy}$ will have to wait at stations thus reducing the passenger-related costs).

Constraint \eqref{eq2} ensures that we will not assign more vehicles than the total number of vehicles available, $N$, constraint \eqref{eq3} ensures that the selected frequency is not higher than the maximum allowed frequency, constraint \eqref{eq4} ensures that the service frequency is lower than the number of vehicles divided by the round-trip travel time and constraint \eqref{eq6} that the determined frequency can only take values greater than 1 vehicle per hour. In the next section, we expand the approach of \citet{furth1981setting} to a network-wide frequency settings problem by allowing the redistribution of vehicles among lines and adding elements to the problem formulation in order to account for the impact of distancing measures.

\section{Network-wide frequency setting model that considers distancing}\label{sec3}

\subsection{Model formulation}
Let us consider a set of lines $L=\{1,...,l,...,|L|\}$ operating in a metro network. Each line $l$ serves a number of metro stations $S_l=\{1,2,...,|S_l|\}$. Let $B_{l,sy}$ be the expected passenger demand between station $s\in S_l\setminus\{|S_l|\}$ and station $y\in S_l~|~y>s$ of line $l$ within an 1-hour period (corresponding to the passenger arrival rate). Before introducing our nomenclature, we list the main assumptions of our study:

\begin{itemize}
    \item[(1)] The passenger arrival rate $B_{l,sy}$ is stable within each 1-hour period of the day. That is, passenger arrivals at stations are uniformly distributed within the 1-hour period (see \citet{furth1981setting,fu2002design}).
    \item[(2)] The obtained passenger demand patterns from historical data are inelastic to the changes made in service frequencies. 
\end{itemize}

\iffalse
\begin{figure}[H]
    \centering
    \includegraphics[scale=0.7]{Figures/fig1.pdf}
    \caption{Left: metro network considered in this study where passengers can travel between metro stations of one line only by using trains of this line. Right: metro network which is not considered in this study because metro lines 2 and 3 both serve the same corridor connecting metro stations 2 and 3.}
    \label{fig1}
\end{figure}
\fi

The detailed nomenclature of our network-wide frequency setting model that considers distancing measures is presented in Table \ref{table1}.

\begin{table}[H]
\caption{Nomenclature \label{table1}}
\centering
\small
\begin{tabular}{p{3.2cm}p{10.0cm}}
\textit{Sets}\\\\
$L=\langle 1,...,l,...|L|\rangle$ &ordered set of metro lines (note that a bi-directional line is considered as a single line that continues its service in the opposite direction)\\
$S_l=\langle 1,2,...,|S_l|\rangle$ &ordered set of metro stations served by line $l\in L$\\
$A=\langle 1,2,...,|A|\rangle$ & set of arcs in the metro network system. An arc connects successive stations and can be served by more than one line if they share the same track corridor\\\\
\textit{Indices}\\\\
$l$ &train line\\
$s$ &metro station\\\\
\textit{Parameters}\\\\
$B_{l,sy}$ &expected hourly passenger arrival rate at station $s$ for passengers whose destination is $y$ and are willing to use line $l$\\
$T_l$&round-trip travel time of line $l\in L$ considering both directions in case of a bi-directional line\\
$c_l$&capacity of trains operating in line $l$\\
$k_l$& maximum passenger load inside each train operating in line $l$ to conform with social distancing\\
$W$&cost of deploying an extra train\\
$V$&value of passenger's time\\
$M$& fare price per km traveled\\
$d_{l,sy}$&traveling distance between stations $s$ and $y$ of line $l\in L$\\
$f_{l,max}$&maximum possible line frequency for line $l\in L$ to maintain safe headways among successive trains serving the same arc (e.g., track corridor)\\
$N$&number of available trains that can be distributed among all lines\\\\
\multicolumn{2}{l}{\textit{Variables}}\\\\
$x_l$&number of trains assigned to line $l\in L$\\
$f_l$& hourly frequency of line $l\in L$\\
$h_l$& time headway among successive trains of line $l\in L$\\
$\gamma_{l,s}$& train load of each train serving line $l$ when it departs from station $s$\\
$b_{l,sy}$& hourly passenger demand between stations $s$ and $y$ of line $l$ that can be served by the trains of line $l$ while conforming to distancing requirements\\
$\tilde{b}_{l,sy}$& hourly passenger demand between stations $s$ and $y$ of line $l$ that cannot be accommodated by the metro services due to distancing requirements\\

\end{tabular}
\end{table}

Let $x_l$ be the number of trains assigned to each line $l\in L$ within a certain time period. Let also $T_l$ be the round-trip travel time of line $l$. Then, the hourly frequency of line $l$ should satisfy the inequality constraint \eqref{eq_f_l} because the service frequency is limited by the number of assigned vehicles to line $l$:

\begin{equation}\label{eq_f_l}
f_l\leq \frac{x_l}{T_l} ~~~~(~\forall l\in L~)
\end{equation} 

In addition, the service frequencies of all lines $l\in L$ traversing arc $a\in A$ should not exceed a maximum allowed frequency to ensure safe headways between trains operating in the same corridor (e.g., we cannot allow trains operating too close to each other because tracks are split into blocks into which only one train can enter at a time due to traffic safety constraints). That is,

\begin{equation}\label{tracks}
  f_l\leq f_{l,max}  ~~~~(~\forall l\in L~)
\end{equation}

Next, let $\gamma_{l,s}$ be the average train load of each train serving line $l$ when it departs from station $s$. We hereby assume that:
\begin{itemize}
\item[(1)] passengers within each train try to maintain the maximum possible distance between each other,
\item[(2)] and all trains serving line $l$ have the same capacity, $c_l$. 
\end{itemize}

This implies a new corona-era capacity limit, $k_l$, where $k_l<c_l$ is the maximum train load for trains of line $l$ below which all passengers can keep a safe distance with each other. This corona-era capacity limit should not be exceeded at any station to ensure that the risk of COVID-19 infections is minimized. This can be expressed by the following inequality constraint:

\begin{equation}
\gamma_{l,s}\leq k_l~~~~(~\forall l\in L,~\forall s\in S_l~)
\end{equation}

If $b_{l,sy}\leq B_{l,sy}$ is the actual passenger demand between stations $s$ and $y$ that can be accommodated by the allocated trains to line $l$ without exceeding the corona-era capacity limit, then 

\begin{equation}
b_{l,sy}=B_{l,sy}-\tilde{b}_{l,sy}~~~~(~\forall l\in L,~\forall s\in S_l,~\forall y\in S_l~|~y\geq s~)
\end{equation}

where $\tilde{b}_{l,sy}$ is the unaccommodated passenger demand that should be served outside our metro system. The passenger demand that is not accommodated by the metro system represents a loss. When refusing to accommodate demand $\tilde{b}_{l,sy}$, the public transport service provider incurs a loss. From a societal perspective this loss can mean reduced accessibility. From the service provider perspective this may mean offering a compensation, arranging alternative means of transport, or simply the revenue loss of ticket sales. We hereby assume that the loss is proportional to the travel distance and we adopt the distance-based fare as a proxy of the loss per unserved passenger-km. If $d_{l,sy}$ is the distance between stations $s$ and $y$ of metro line $l$, the operator will lose a revenue of $M\cdot \tilde{b}_{l,sy}\cdot d_{l,sy}$ because $\tilde{b}_{l,sy}$ passengers are refused service and $M$ is the average ticket fee per km traveled. Then, our objective function for the network-wide frequency setting problem that considers social distancing becomes:

\begin{equation}
z(x,f,b,\tilde b):= W\sum_{l\in L}x_l + V\sum_{l\in L}\sum_{s\in S_l\setminus\{|S_l|\}}\sum_{y\in S_l~|~y>s} b_{l,sy}\frac{1}{f_l} + \sum_{l\in L}\sum_{s\in S_l\setminus\{|S_l|\}}\sum_{y\in S_l~|~y>s} M d_{l,sy} \tilde{b}_{l,sy}
\end{equation}

The first term is the cost of operating the vehicles, the second term is the cost related to passengers' waiting times and the third term is the cost of the revenue losses associated with the passengers that are refused service. The objective function is formulated as a compensatory monetary term and expresses a generalized cost. Considering the above formulation, the mathematical program for the network-wide frequency setting problem that considers distancing measures is formulated as follows:

\begin{align}
(\text{Q})\quad & \min_{x,f,b,\tilde b} & z(x,f,b,\tilde b)\label{eq10}\\
&\text{s.t.} & \sum_{l\in L}x_l\leq N&\label{eq11}\\
&& f_l\leq \frac{x_l}{T_l}&~~(~\forall l\in L~)\label{eq12}\\
&&  f_l\leq f_{l,max}  &~~(~\forall l\in L~)\label{eq13}\\
&& \gamma_{l,s}\leq k_l&~~(~\forall l\in L,~\forall s\in S_l~)\label{eq14}\\
&& \gamma_{l,1}=\sum_{y\in S_l}b_{l,1y}\frac{1}{f_l}&~~(~\forall l\in L~)\label{eq15}\\
&& \gamma_{l,s}=\gamma_{l,s-1}-\sum_{y\in S_l~|~y<s}b_{l,ys}\frac{1}{f_l}+\sum_{y\in S_l~|~y>s}b_{l,sy}\frac{1}{f_l}&~~(~\forall l\in L,~\forall s\in S_l\setminus\{1\}~)\label{eq16}\\
&& b_{l,sy}=B_{l,sy}-\tilde{b}_{l,sy}&~~(~\forall l\in L,~\forall s\in S_l,~\forall y\in S_l~|~y\geq s~)\label{eq17}\\
&& x_l\in\mathbb{Z}_{\geq 0}&~~(~\forall l\in L~)\label{eq18}\\
&& f_l\in \mathbb{R}_{\geq 1}&~~(~\forall l\in L~)\label{eq19}
\end{align}
 
Note that constraint \eqref{eq11} ensures that the number of all assigned trains within our time period does not exceed $N$. In addition, constraint \eqref{eq15} determines the train load when a train departs the terminal as the number of passengers that board at station 1 and will alight at any other station $y$ within our 1-hour time period, $b_{l,sy}$, divided by the hourly frequency, $f_l$. Similarly, the recursive equation \eqref{eq16} determines the train load when a train belonging to line $l$ departs from station $s\in S_l \setminus\{1\}$. This is a passenger flow conservation equation where the train load is equal to the train load when departing from the previous station, $\gamma_{l,s-1}$, minus the number of passengers that alight at station $s$, $\sum\limits_{y\in S_l~|~y<s}b_{l,ys}\frac{1}{f_l}$, plus the number of passengers that board at station $s$ and will alight at any other station $y>s$ of line $l$.

The mathematical program for the network-wide frequency setting problem that considers distancing measures, $(Q)$, is a mixed-integer nonlinear programming problem (MINLP) because of its fractional objective function and its fractional constraints \eqref{eq15} and \eqref{eq16}.

\subsection{Reformulation to a mixed-integer quadratic program (MIQP)}
The MINLP problem $(Q)$ can be transformed to an easier-to-solve mixed-integer quadratic program (MIQP) that neither contains a fractional objective function nor fractional constraints. Let us consider a line headway of $h_l=\frac{1}{f_l},~\forall l\in L$. Then, the fractional constraint in Eq.\eqref{eq15} becomes $\gamma_{l,1}=\sum_{y\in S_l}b_{l,1y}h_l~~(~\forall l\in L~)$. In addition, the fractional constraint \eqref{eq16} becomes $\gamma_{l,s}=\gamma_{l,s-1}-\sum_{y\in S_l~|~y<s}b_{l,ys}h_l+\sum_{y\in S_l~|~y>s}b_{l,sy}h_l~~(~\forall l\in L,~\forall s\in S_l\setminus\{1\}~)$.

Constraint \eqref{eq12} is transformed into the following quadratic constraint: $h_lx_l\geq T_l~~(~\forall l\in L~)$. Finally, constraint \eqref{eq13} becomes $h_l\geq \frac{1}{f_{l,max}}  ~~(~\forall l\in L~)$ and the objective function takes the following form:

\begin{equation}
z(x,h,b,\tilde b):= W\sum_{l\in L}x_l + V\sum_{l\in L}\sum_{s\in S_l\setminus\{|S_l|\}}\sum_{y\in S_l~|~y>s} b_{l,sy}h_l + \sum_{l\in L}\sum_{s\in S_l\setminus\{|S_l|\}}\sum_{y\in S_l~|~y>s} M d_{l,sy} \tilde{b}_{l,sy}
\end{equation}

This reformulation results in the following MIQP:

\begin{align}
(\tilde{\text{Q}})\quad & \min_{x,h,b,\tilde b} & z(x,h,b,\tilde b)\label{eq23}\\
&\text{s.t.} & \sum_{l\in L}x_l\leq N&\label{eq24}\\
&& h_lx_l\geq T_l&~~(~\forall l\in L~)\label{eq25}\\
&&  h_l\geq \frac{1}{f_{l,max}}  &~~(~\forall l\in L~)\label{eq26}\\
&& \gamma_{l,s}\leq k_l&~~(~\forall l\in L,~\forall s\in S_l~)\label{eq27}\\
&& \gamma_{l,1}=\sum_{y\in S_l}b_{l,1y}h_l&~~(~\forall l\in L~)\label{eq28}\\
&& \gamma_{l,s}=\gamma_{l,s-1}-\sum_{y\in S_l~|~y<s}b_{l,ys}h_l+\sum_{y\in S_l~|~y>s}b_{l,sy}h_l&~~(~\forall l\in L,~\forall s\in S_l\setminus\{1\}~)\label{eq29}\\
&& b_{l,sy}=B_{l,sy}-\tilde{b}_{l,sy}&~~(~\forall l\in L,~\forall s\in S_l,~\forall y\in S_l~|~y\geq s~)\label{eq29}\\
&& x_l\in\mathbb{Z}_{\geq 0}&~~(~\forall l\in L~)\label{eq30}\\
&& h_l\in \mathbb{R}_{\geq 0}&~~(~\forall l\in L~)\label{eq31}
\end{align}

\section{Model application}\label{sec4}
\subsection{Case study description}
\begin{figure}[H]
\centering
\includegraphics[scale=0.6]{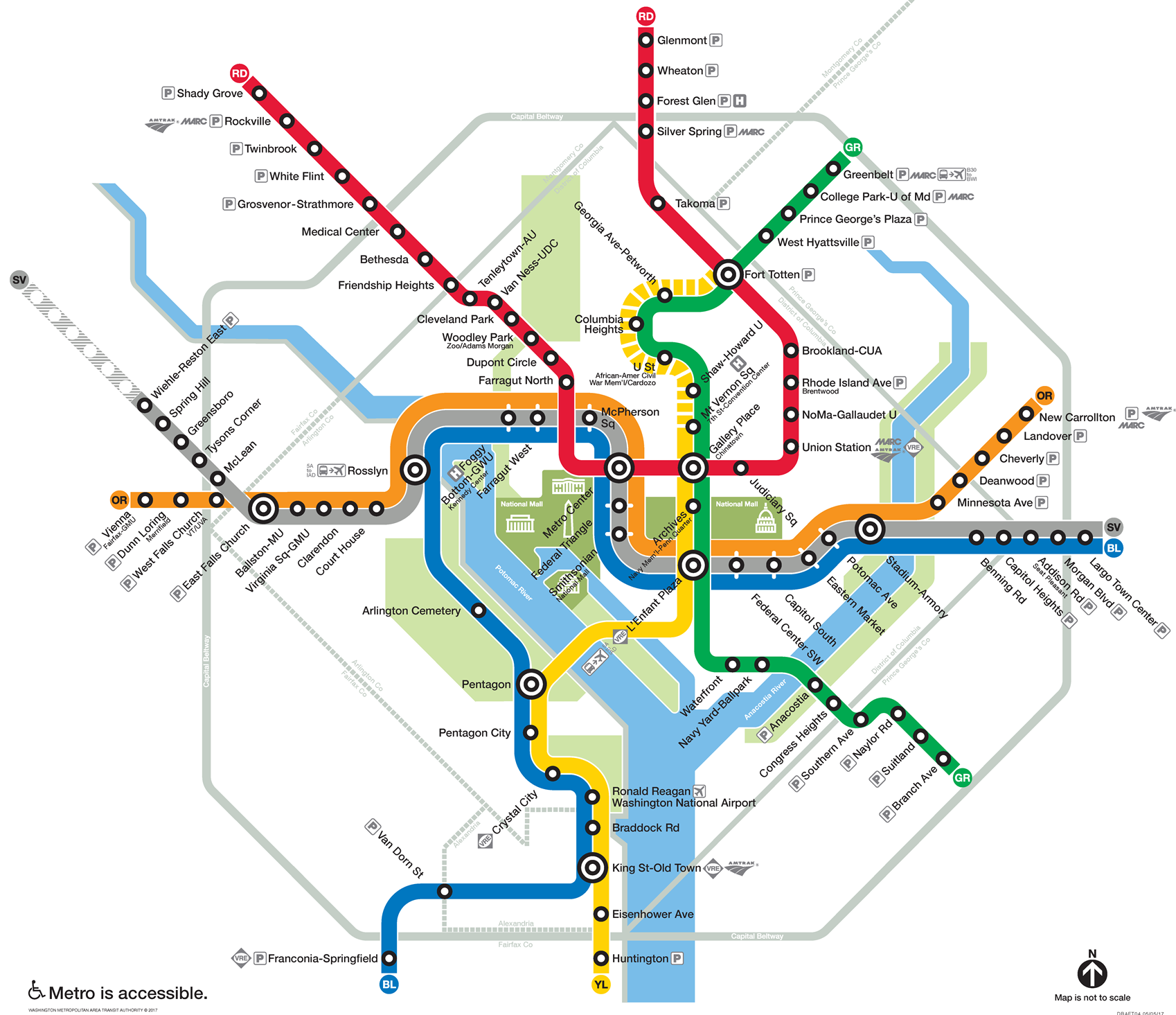}
\caption{Washington D.C. metro network map, source: \url{www.wmata.com}\label{fig2}}
\end{figure}

We apply our model to the case study of the Washington D.C. Metro system of the Washington Metropolitan Area Transit Authority (WMATA). It is a rapid transit system serving the Washington metropolitan area in the United States. The network consists of six lines, ninety-one metro stations. The metro system serves the District of Columbia, as well as several jurisdictions in the states of Maryland and Virginia. It is the second busiest rapid transit system in the United States with 295 million passenger trips in 2018 \citep{wmatafacts}. In 2019, the average weekday ridership was 626 thousand trips \citep{wmatafacts2019}. Passengers validate their smart card (SmartTrip) upon entering an origin station and leaving a destination station, and the fee is determined based on the traveled distance of the origin-destination combination. The network configuration consisting of the six metro lines is presented in Fig.\ref{fig2}.

Due to infrastructure capacity limits, the maximum allowed frequency per track segment is 30 vehicles per hour, resulting in a minimum permitted headway of 2 minutes at common track corridors. From Fig.\ref{fig2} one can note that there are several track corridors that are traversed by vehicles serving more than one line. For the purpose of our case study, we assume the availability of $N=140$ trains. Table \ref{tab:distances} provides information on each of the service lines, including the number of stations, total bi-directional track length, and the names of the terminal stations. 

\begin{table}[H]
  \centering
  \footnotesize
  \caption{Terminals and line distances}
    \begin{tabular}{lrrrr}
    \toprule
          &       &       & \multicolumn{2}{c}{Terminals} \\
    \cmidrule{4-5}
          & stations  & distance in km & Western/Southern & \multicolumn{1}{c}{Eastern/Northern} \\
        &(per direction)&(both directions)& \\\cmidrule{2-5}
    orange line & 26    & 85    & Vienna & New Carrollton \\
    blue line & 27    & 97.6  & Franconia–Springfield & Largo Town Center \\
    silver line & 28    & 95.2  & Wiehle–Reston East & Largo Town Center \\
    green line & 21    & 74.16 & Branch Avenue & Greenbelt \\
    red line  & 26    & 102.6 & Shady Grove & Glenmont \\
    yellow line& 22    & 48.5  & Huntington & Greenbelt \\
    \bottomrule
    \end{tabular}%
  \label{tab:distances}%
\end{table}%

\begin{figure}[H]
\centering
\includegraphics[scale=0.9]{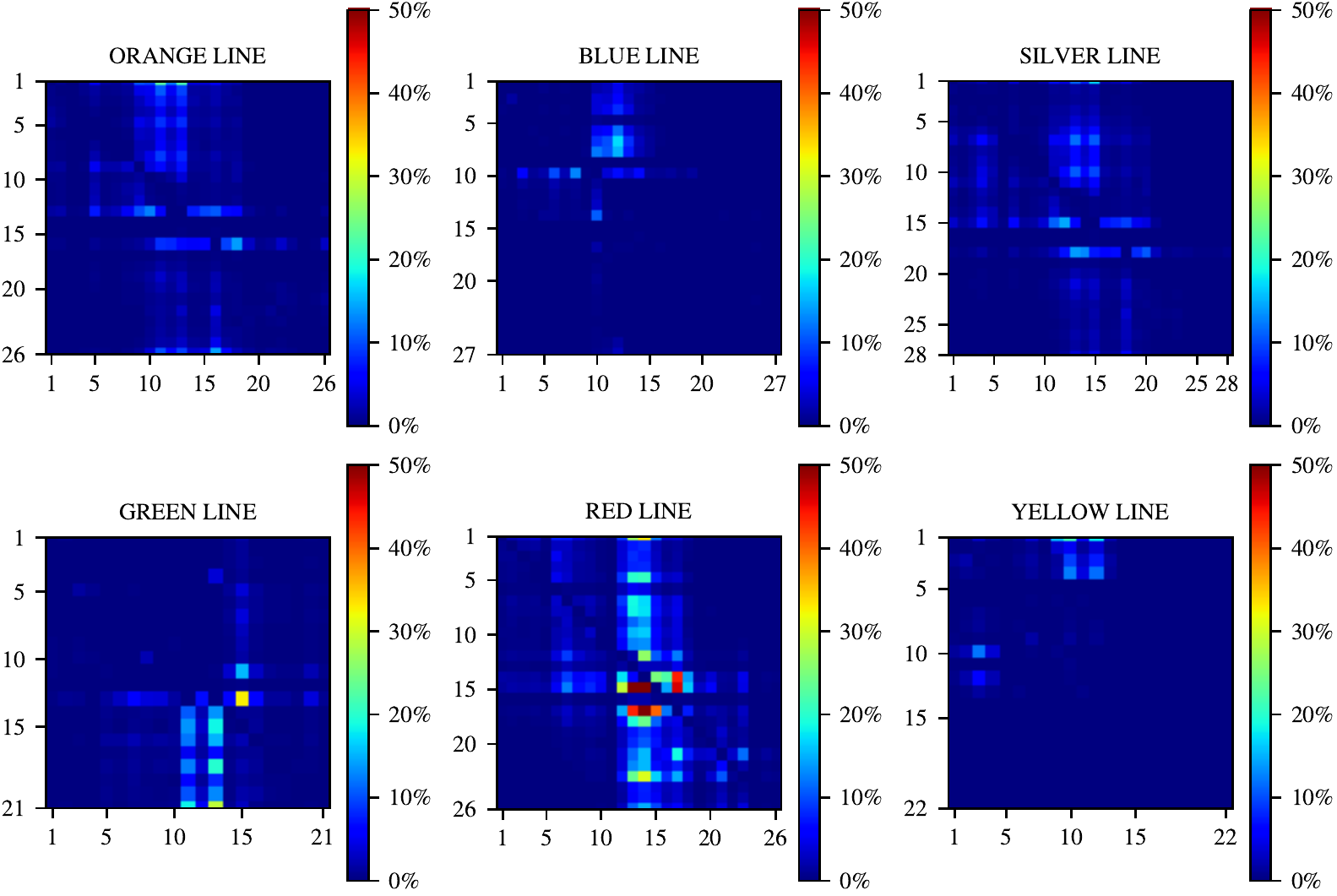}
\caption{Hourly passenger demand per origin-destination pair for each line during the peak hour of 8-9am as a percentage of the actual vehicle capacity\label{fig:demand}} 
\end{figure}

In order to investigate the impact of distancing measures, we concentrate on the rush hour (8:00-9:00 am) of a typical weekday. In particular, we examine the need to deploy additional trains and possibly refuse boardings. The hourly rate of passengers arriving at station $s$ of line $l$ and destined to station $y$ is presented in Fig.\ref{fig:demand} as a percentage of the vehicle capacity. This expected hourly arrival rate is derived from the analysis of historical smart card data with complete information regarding passenger boarding and alighting locations from 20 working days in September 2018. The line-specific origin-destination matrices have been inferred using the origin, destination, and transfer inference (ODX) method as detailed in \citep{sanchez2017inference}.

\begin{figure}[H]
    \centering
    \includegraphics[scale=1.0]{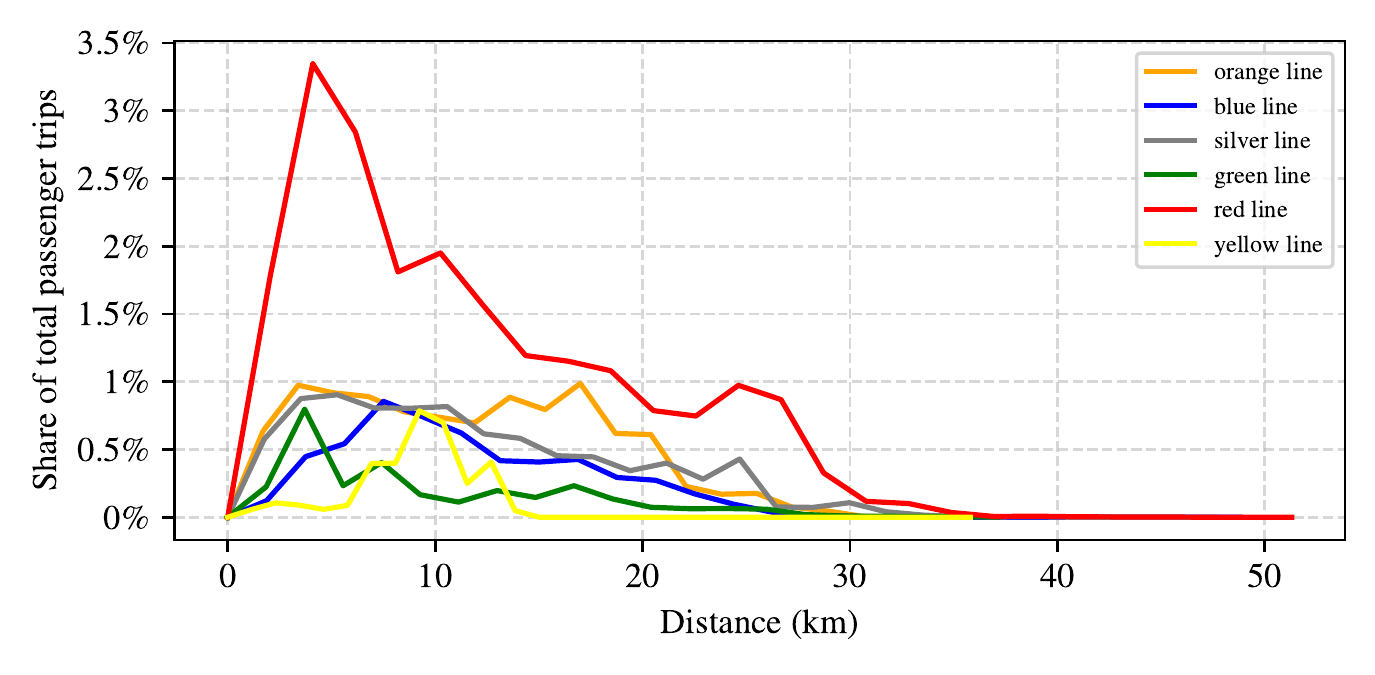}
    \caption{Share of passenger trips of each metro line from 8am to 9am with respect to the traveled distance }
    \label{fig:demkm}
\end{figure}

The distances travelled by each of the metro lines vary considerably. Fig.\ref{fig:demkm} shows the share of passenger trips of a certain traveled distance for each of the lines. This share is calculated as the number of passenger trips of a line for a specific travel distance divided by the total number of trips performed by all lines. As can be seen in Fig.\ref{fig:demkm} the Red line has up to three times more passenger trips than other lines and passengers trips on the Red, Orange and Silver lines travel longer distances than those on the Blue, Yellow and Green lines.

The values of the remaining model parameters are given in Table \ref{tab:parameters}. Each metro train consists of 6 to 8 rail cars, with an average maximum capacity of 1700 people. During the morning period (5am-12pm) a metro train carries an average of 133 travelers at any given time. This value, however, varies greatly with some central segments reaching capacity limits. Assuming that passengers are evenly spaced across platforms, each train operating in any line $l\in L$ can carry only $k_l=703$ people at any given time to satisfy a 1-meter distancing requirement, $k_l=312$ to fulfill a 1.5-meter distance, and $k_l=176$ to comply with a 2-meter distance (see \citet{krishnakumari2020virus}).

\begin{table}[H]
  \centering
  \footnotesize
  \caption{Parameter values}
    \begin{tabular}{lrrrrrr}
    \toprule
          & \multicolumn{6}{c}{lines} \\
    \cmidrule{2-7}
          & orange & blue  & silver & green & red   & yellow \\\midrule
    $T_l$ (minutes)  & 150   & 156   & 170   & 122   & 160   & 120 \\
    $c_l$ (passengers per train)  & 1700  & 1700  & 1700  & 1700  & 1700  & 1700 \\
    $k_l$ without social distancing & 1700  & 1700  & 1700  & 1700  & 1700  & 1700 \\
    $k_l$ with 1 meter social distancing & 703   & 703   & 703   & 703   & 703   & 703 \\
    $k_l$ with 1.5 meter social distancing & 312   & 312   & 312   & 312   & 312   & 312 \\
    $k_l$ with 2 meters social distancing & 176   & 176   & 176   & 176   & 176   & 176 \\
    \bottomrule
    \end{tabular}%
  \label{tab:parameters}%
\end{table}%

Passengers' value of time in our case study is $V=14.67 \textdollar$ per hour, or, equivalently, 24.4 cents per minute (based on the value of travel time reported in \citet{white2016revised} and adjusting the value by 1\% per year). In addition, we have two parameters $W$ and $M$ corresponding to the operational cost of using an extra train and the cost associated with the travel distances of passengers who are refused service, respectively. The value of parameter $W$ depends on the preferences of the metro operator since some operators might be willing to use all their available trains to reduce the passenger waiting times, whereas other operators might seek an economically beneficial trade-off between operational costs and passenger waiting times. We tested our model with different values of $W$ and selected a cost of $W=36.675\textdollar$ per extra train since this cost returns the same train allocation to lines when solving program $(\tilde Q)$ for the case of no distancing (i.e. normal capacity levels) as the planned train allocation actually performed by the metro operator. Further, the passenger fare at the metro system depends on the travel distance and the time period (peak/off-peak periods). The 8am-9am is a peak hour and the fare cost per passenger is found to have an average value of $M=0.7\textdollar$ per km traveled.

\subsection{Scenario design and implementation}
The topic of distancing is currently contested as the transmission of COVID-19 is not yet well characterized. It is likely to be similar to SARS, which was spread by contact, droplet, and airborne routes \citep{yu2004evidence}. The World Health Organization and most countries recommend an 1.5- to 2-meter social distancing. In a study with 94 patients of influenza, \citet{bischoff2013exposure} showed that the virus can be transmitted up to $\sim$1.9 meters from patients during non–aerosol-generating patient-care activities. It is important to note that increasing the distance among passengers does not proportionally decrease the probability of transmitting COVID-19. 

To assess the impact of distancing measures, we consider the following four scenarios:

\begin{itemize}
\item[(I)] the do-nothing scenario that does not consider distancing measures and can utilize the full capacity of 1700 passengers per train;
\item[(II)] the 1-meter distancing scenario that allows train loads of up to 703 passengers per train;
\item[(III)] the 1.5-meter distancing scenario that allows train loads of up to 312 passengers per train;
\item[(IV)] the 2-meter distancing scenario that allows train loads of up to 176 passengers per train.
\end{itemize}

We program and solve our MIQP model in program $(\tilde Q)$ using LINGO 18.0. LINGO uses the LINDO software package for nonlinear integer optimization \citep{schrage1986linear}. Our numerical experiments are executed in a general-purpose computer with Intel Core i7-7700HQ CPU @ 2.80GHz and 16 GB RAM. In all four scenarios, LINGO was capable of returning a locally optimal solution within less than 20 seconds using branch and bound and quadratic programming. To facilitate the reproduction of our model to other networks, its source code is publicly released at \citet{gkiotsalitis2020covid}. 

\subsection{Results and analysis}
The objective function score(s) of the solution of each scenario as well as the computation times are presented in Table \ref{tab:computation}.

% Table generated by Excel2LaTeX from sheet 'Overall_lines'
\begin{table}[H]
  \centering
  \footnotesize
  \caption{Objective function value of the respective solution, iterations until convergence, and computational costs}
    \begin{tabular}{lrrr}
    \toprule
          & Objective function score (\$) & iterations until convergence & computation time (sec) \\
    \midrule
    I. Do-nothing & 5980.5 & 626   & 17.08 \\
    II. 1-meter & 6362.3 & 259   & 15.78 \\
    III. 1.5-meter & 85390.5 & 747   & 16.33 \\
    IV. 2-meter & 210835.9 & 244   & 14.95 \\
    \bottomrule
    \end{tabular}%
  \label{tab:computation}%
\end{table}%

The decision variable values in the optimal solutions under each distancing scenario are reported in Table \ref{tab:allocation} and present the allocated vehicles per line and the resulting service headway.

\begin{table}[H]
  \centering
  \footnotesize
  \caption{Assigned trains, $x_l$, and headway (in minutes), $h_l$, per line $l\in L$ at each one of the social distancing scenarios}
    \begin{tabular}{lrrrrrrrrrrrr}
    \toprule
          &       & \multicolumn{11}{c}{social distancing scenarios} \\
    \cmidrule{3-13}
          &       & \multicolumn{2}{c}{I. Do-nothing} & \multicolumn{1}{c}{} & \multicolumn{2}{c}{II. 1-meter} & \multicolumn{1}{c}{} & \multicolumn{2}{c}{III. 1.5-meter} & \multicolumn{1}{c}{} & \multicolumn{2}{c}{IV. 2-meter} \\\cmidrule{3-4}\cmidrule{6-7}\cmidrule{9-10}\cmidrule{12-13}
          &       & trains & headway &       & trains & headway &       & trains & headway &       & trains & headway \\\midrule
    orange &       & 15    & 10    &       & 19    & 8     &       & 23    & 7     &       & 28    & 6 \\
    blue  &       & 10    & 16    &       & 14    & 12    &       & 9     & 18    &       & 6     & 26 \\
    silver &       & 15    & 12    &       & 19    & 9     &       & 23    & 8     &       & 24    & 7 \\
    green &       & 10    & 13    &       & 14    & 9     &       & 15    & 9     &       & 9     & 14 \\
    red   &       & 25    & 7     &       & 41    & 4     &       & 66    & 3     &       & 69    & 3 \\
    yellow &       & 7     & 18    &       & 7     & 18    &       & 4     & 30    &       & 4     & 30 \\\midrule
    total &       & 82    & n/a   &       & 114   & n/a   &       & 140   & n/a   &       & 140   & n/a \\
    \bottomrule
    \end{tabular}
  \label{tab:allocation}%
\end{table}%

From Table \ref{tab:allocation} one can note that at the do-nothing case we only deploy 82 trains out of the $N=140$. The same passenger demand requires the utilization of 114 trains in the case of 1-meter distancing (case II). Finally, scenarios III and IV require the deployment of all available trains. 

\begin{figure}[H]
\centering
\includegraphics[scale=0.9]{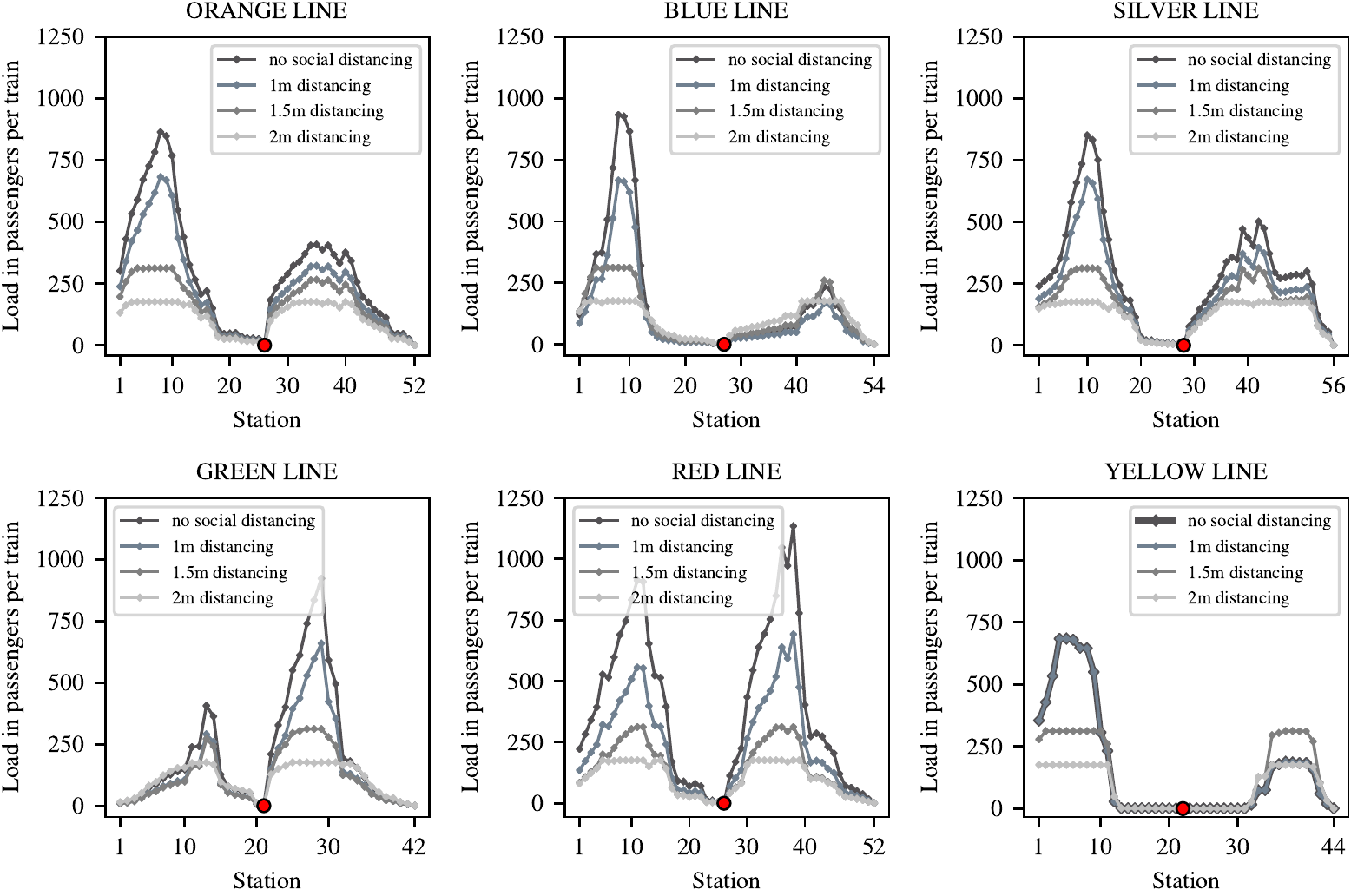}
\caption{Passenger load of each train operating from 8-9am at each metro station in scenarios I, II, III and IV. The red markers indicate the change of line direction. \label{fig:loads}}
\end{figure}

In addition to the increase in operational costs due to the deployment of more vehicles and train drivers, a number of passengers are also refused service in scenarios III and IV. In Fig.\ref{fig:loads} we present the train load at each station for each one of the six lines. The normal vehicle capacity of $c_l=1700$ passengers is not reached at the base case scenario once the fleet is optimally allocated. The respective optimal solution results in train loads that never exceed 1250 passengers. It can also be observed that the optimal allocation results with trains loads not exceeding 703 passengers per train on the Yellow line in the base case (scenario I). Consequently, there is sufficient residual capacity to also accommodate the demand when subject to 1-meter  distancing (scenario II). In contrast, all other lines require re-allocating trains to ensure that the new capacity limit is not violated when switching from scenario I to scenario II, as can be seen in Table \ref{tab:allocation}. 

As specified in the problem constraints, the optimal solution for case II does not allow train loads of more than 703 passengers, for case III of more than 312, and for case IV of more than 176. This results in different train occupancy levels for each distancing scenario. Let the train occupancy be defined as the train load divided by the number of seats (in our case study, each train has $\tilde{c}_l=616$ available seats):

\begin{equation}\label{occline}
    \text{Occupancy}=\frac{1}{|S_l|}\sum_{s_\in S_l}\frac{\gamma_{l,s}}{\tilde{c}_l}~~~~(~\forall~l\in L~)
\end{equation}

The average train occupancy level per line for each one of the distancing scenarios is presented in Table \ref{tab:occupancy}. The occupancy level averaged over all line segments ranges in the optimal solution for the base case from 28\% for the Blue and Yellow lines to 66\% for the Red line. These levels drop to 14\%-19\% for all lines when complying with  2-meter distancing (scenario IV).

\begin{table}[H]
  \centering
  \footnotesize
  \caption{Average train occupancy for each distancing scenario}
    \begin{tabular}{rrrrr}
    \toprule
          & \multicolumn{4}{c}{social distancing scenarios} \\
    \cmidrule{2-5}
          & I. Do-nothing & II. 1-meter & III. 1.5-meter & IV. 2-meter \\\midrule
    orange & 50\%    & 39\%    & 28\%    & 19\% \\
    blue  & 28\%    & 19\%    & 19\%    & 17\% \\
    silver & 47\%    & 39\%    & 28\%    & 19\% \\
    green & 36\%    & 25\%    & 19\%    & 18\% \\
    red   & 66\%    & 39\%    & 22\%    & 17\% \\
    yellow & 28\%    & 28\%    & 22\%    & 14\% \\
    \bottomrule
    \end{tabular}%
  \label{tab:occupancy}%
\end{table}%

\begin{figure}[H]
\centering
\includegraphics[scale=0.9]{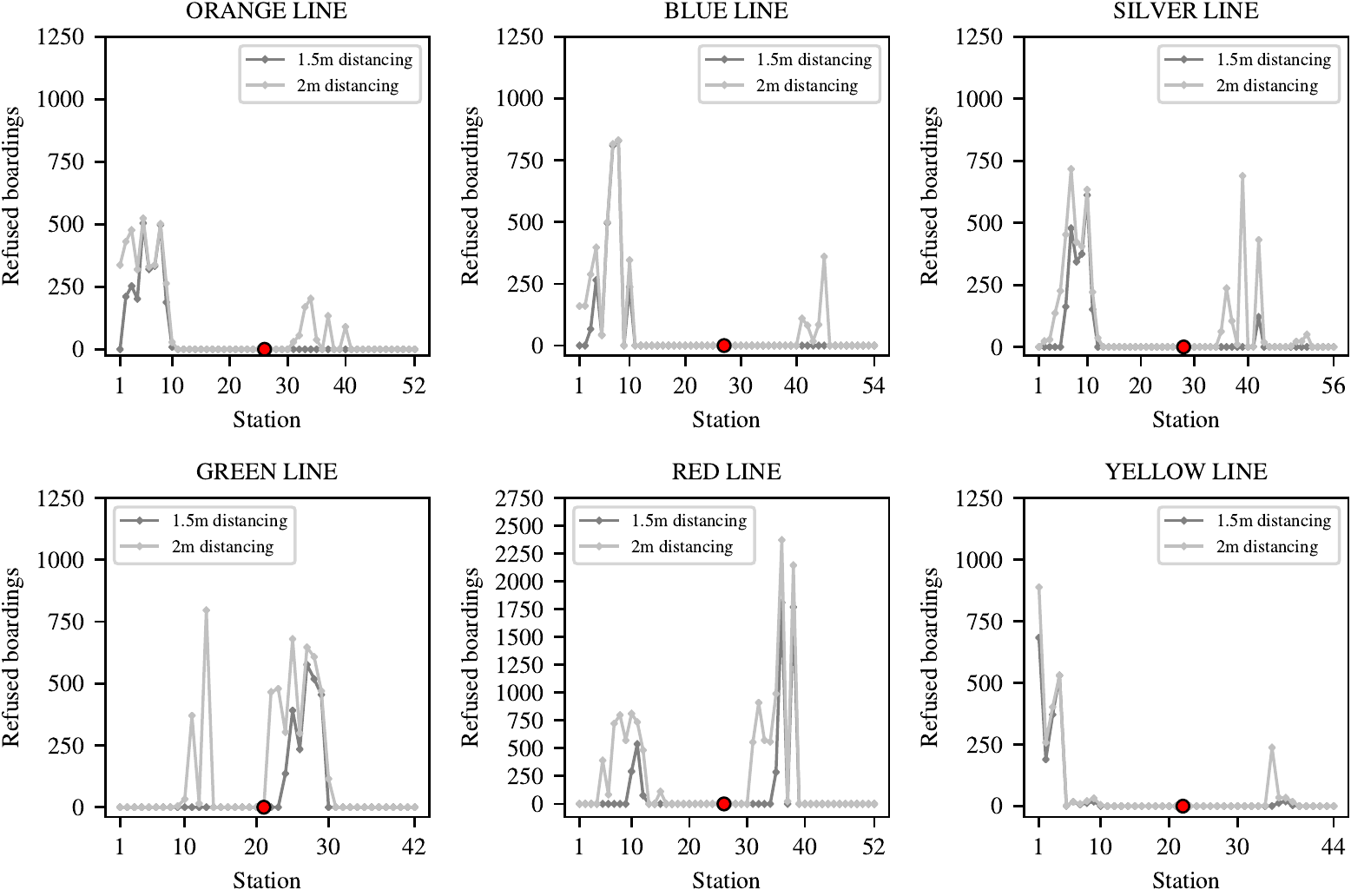}
\caption{Refused passenger boardings per station during the 8-9am period in 1.5- and 2-meter distancing scenarios. The red markers indicate the change of line direction\label{fig:figrefusedb}}
\end{figure}

When enforcing distancing measures of 1.5- or 2-meters, even after optimally reallocating the fleet, some passengers are refused service. In Fig.\ref{fig:figrefusedb} we report the number of refused boardings at each metro station for the 1.5- and 2-meters distancing policies. The Red line which has the highest passenger demand also sees the highest number of refused passenger boardings. This happens in spite of our optimization model assigning 66 and 69 trains to the Red line in the cases of 1.5- and 2-meter distancing, respectively (see Table \ref{tab:allocation}). Note that allocating more trains to the Red line (i.e. 70 trains) would have resulted in violating the maximum frequency constraint that requires a minimum headway of 2 minutes.

To summarize the results of the impact of considering social distancing in the planning phase of metro operations, we use the following key performance indicators:

\begin{itemize}
    \item[(1)] the average value of the maximum possible passenger waiting times, $$\mathcal{O}_1:= \sum\limits_{l\in L}\sum\limits_{s\in S_l}\sum\limits_{y\in S_l~|~y\geq s} b_{l,sy}\frac{1}{f_l}$$    
    \item[(2)] the number of trains that are needed to operate the metro service, which indicate the operational costs, $\mathcal{O}_2:=\sum\limits_{l\in L}x_l$ 
    \item[(3)] the average train occupancy, $\mathcal{O}_3~(\%)$ 
    \item[(4)] the number of instances where passengers cannot maintain a distance of at least 1.5 meter from each other, $\mathcal{O}_4:=\sum\limits_{l\in L}\sum\limits_{s\in S_l} r_{l,s}$ where $r_{l,s}=0$ if $\gamma_{l,s}\leq 312$, and $r_{l,s}=1$ otherwise.
    \item[(5)] the number of passengers who are refused service multiplied by their traveled distance, $\mathcal{O}_5:=\sum\limits_{l\in L}\sum\limits_{s\in S_l}\sum\limits_{y\in S_l~|~y>s} \tilde b_{l,sy}\cdot d_{l,sy}$
    \item[(6)] the number of passengers who are refused service, $\mathcal{O}_6:=\sum\limits_{l\in L}\sum\limits_{s\in S_l}\sum\limits_{y\in S_l~|~y>s} \tilde b_{l,sy}$
\end{itemize}

\begin{figure}[H]
\centering
\includegraphics[scale=1.0]{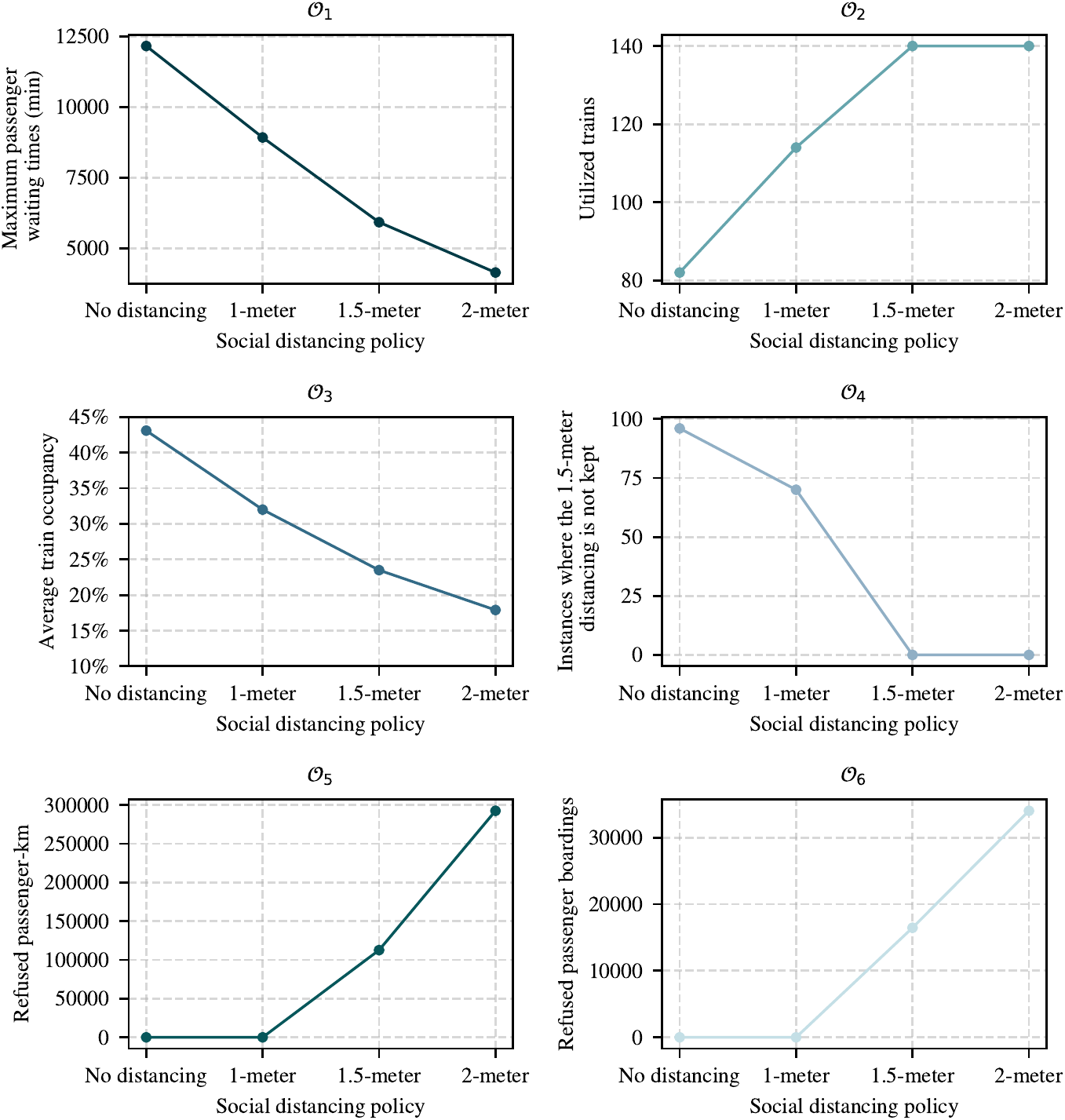}
\caption{Values of key performance indicators under different social distancing scenarios \label{fig:overall}}
\end{figure}

\section{Discussion and conclusion}\label{sec5}
We propose a network-wide model that can set the optimal frequencies of services lines under different distancing scenarios and apply it to the case study network of the Washington D.C. metro. The model determines the optimal fleet allocation while considering unsatisfied demand in the event that not all demand can be absorbed due to the varying capacity limits of different distancing policies.

The results of our model indicate that the normal capacity limit is not binding in the optimal solution for the base case scenario with no distancing requirements. Moreover, all passenger demand can be satisfied when deploying 82 out of the 140 trains available so as to optimally balance between the operational costs and the passenger waiting times at stations. Furthermore, it is possible to comply with the 1m distancing requirements while still accommodating all passengers by increasing the service frequency, hence requiring a larger fleet size of 114 vehicles. This also yields shorter passenger waiting times. This does imply, however, that in about 25\% of the instances, a distancing of less than 1.5m occurs, potentially inducing public health risks (see the graph of the key performance indicator $\mathcal{O}_4$ in Fig.\ref{fig:overall}).

Stricter distancing policies come at a greater cost. Enforcing 1.5m distancing requires deploying the maximum number of trains assumed available in our case study. Even though the entire fleet of 140 vehicles is utilized, 16,456 passenger-trips of more than 112,582 passenger-km cannot be accommodated. These numbers increase two-fold (34,025 pass trips) and almost three-fold (292,412 pass-km), respectively, when imposing 2m distancing. In the latter case, the average train occupancy considering only the seated capacity plummets below 20\% for all metro lines (see Table \ref{tab:occupancy}). 

As can be expected, our model allocates as many vehicles as possible to the most heavily-utilized line (i.e. Red line) given the safety-related 2-min minimum headway constraint. In contrast, for some other lines (e.g. Yellow), there are more allocated vehicles in the base case scenario with no distancing measures than in the scenarios with 1.5m and 2m distancing where 58 more vehicles are deployed (see Table \ref{tab:allocation}). This is an interesting finding since one might have expected that the number of trains assigned to each line will increase (or at least will not decrease) when allocating an overall larger fleet and imposing stricter capacity constraints. However, when the permitted vehicle capacity is enough to satisfy the passenger demand (i.e. in the base case scenario), the relatively lightly-utilized lines receive more vehicles to reduce passenger waiting times at stations since this is one of our main objectives. Notwithstanding, when the permitted capacity is reduced to comply with the distancing policies and we cannot anymore accommodate all passengers despite having employed all available resources, then vehicles are redistributed from lightly-utilized lines to heavily-utilized lines that already have to refuse many passengers (see Fig.\ref{fig:demkm}). That is, the inability to accommodate the passenger demand in some lines results in the redistribution of vehicles because of the higher priority given to reducing the number of refused passengers compared to shortening waiting times.

Our model formulation strives for minimizing refused pass-km so as to minimize the associated loss revenues or costs inflicted. The reallocation of trains over lines under stricter distancing policies is hence not only driven by the wish to reduce the number of denied boardings, but more specifically the objective to reduce unsatisfied passenger-km. This implies that the model seeks to prioritize the allocation of resources to lines characterized by longer passenger trips over lines that are mostly used for shorter trips. This effect is visible in Table \ref{tab:allocation} as vehicles are reallocated from lines characterized by short passenger trips (e.g. Green, majority of the trips are less than 10km, see Fig. \ref{fig:demkm}) to lines characterized by longer passenger trips (e.g. Orange, many trips are between 10 and 20km) when switching from 1.5m to 2m distancing. This change in resource allocation is hence driven by the underlying demand patterns - passenger volumes and travel distance.

The aforementioned findings of our network-wide frequency setting approach that considers distancing measures can support planning decisions made by public transport service providers in the phasing of exit strategies and the aftermath of the corona crisis. Model formulation strives to balance between operational, passenger, revenue loss, and health-risk related considerations. Depending on the local circumstances, planners and policymakers may assign different terms to the different objectives and compare the resulting performance under different distancing requirements. The model can also be used to dimension service supply for different demand levels during lock-down periods and potential changes associated with the outbreak and evolution of an epidemic. 

Primary directions for further research pertain to refining and modeling the impact of refused demand.
The cost associated with unsatisfied demand can be modified to account for aspects extending beyond revenue losses, such as reduced access to activities, customer retention effect, externalities caused by switching to using a car, or the cost of offering compensation or an alternative mean of transport. Future analysis may consider the equity implications of alternative solutions. The potential of partnerships with on-demand service providers to cater for the unsatisfied demand that exceeds the permitted capacity limits by offering them a (shared)ride as well as the operations of such services can possibly be the subject of further investigation.   

\section{Acknowledgements}
The authors thank Washington
Metropolitan Area Transit Authority and in particular Jordan Holt for their
invaluable cooperation and for providing the data that made this study possible.

\bibliographystyle{tfcad}
\bibliography{interactcadsample}

\end{document}